\documentclass[aip,graphicx,preprint]{revtex4-1}

\usepackage{graphicx}% Include figure files
\usepackage{dcolumn}% Align table columns on decimal point
\usepackage{bm}
\usepackage{upgreek}

\begin{document}

%\preprint{}


\title{Enhanced Photodetection in Graphene-Integrated Photonic Crystal Cavity}% Force line breaks with \\

\author{Ren-Jye Shiue}%
\thanks{These authors contribute equally to this work}
\affiliation{Department of Electrical Engineering and Computer Science, Massachusetts Institute of Technology, Cambridge, Massachusetts 02139, United States}

%\email{tedshiue@mit.edu}
 
\author{Xuetao Gan}%
\thanks{These authors contribute equally to this work} 
\affiliation{Department of Electrical Engineering, Columbia University, New York, New York 10027, United States}%

\author{Yuanda Gao}
\thanks{These authors contribute equally to this work}
\affiliation{Department of Mechanical Engineering, Columbia University, New York, New York 10027, United States}

\author{Luozhou Li}
\affiliation{Department of Electrical Engineering, Columbia University, New York, New York 10027, United States}%

\author{Xinwen Yao}
\affiliation{Department of Electrical Engineering, Columbia University, New York, New York 10027, United States}%

\author{Attila Szep}
\affiliation{Air Force Research Laboratory, Sensors Directorate, WPAFB, Dayton, Ohio 45433, Unite States}

\author{Dennis Walker, Jr.}
\affiliation{Air Force Research Laboratory, Sensors Directorate, WPAFB, Dayton, Ohio 45433, Unite States}

\author{James Hone}
\affiliation{Department of Mechanical Engineering, Columbia University, New York, New York 10027, United States}

\author{Dirk Englund}
\affiliation{Department of Electrical Engineering and Computer Science, Massachusetts Institute of Technology, Cambridge, Massachusetts 02139, United States}
\email{englund@mit.edu}



%\thanks{}


\date{\today}

\begin{abstract}
We demonstrate the controlled enhancement of photoresponsivity in a graphene photodetector by coupling to slow light modes in a long photonic crystal linear defect cavity. Near the Brillouin zone (BZ) boundary, spectral coupling of multiple cavity modes results in broad-band photocurrent enhancement from 1530 nm to 1540 nm. Away from the BZ boundary, individual cavity resonances enhance the photocurrent eight-fold in narrow resonant peaks. Optimization of the photocurrent via critical coupling of the incident field with the graphene-cavity system is discussed. The enhanced photocurrent demonstrates the feasibility of a wavelength-scale graphene photodetector for efficient photodetection with high spectral selectivity and broadband response.

%Valid PACS numbers may be entered using the \verb+\pacs{#1}+ command.

\end{abstract}

%Use showkeys class option if keyword

%\keywords{}

\maketitle

The unique properties of graphene have generated strong interest in developing opto-electronics devices based on the material \cite{Bonaccorso2010,Bao2012}. Examples include graphene-based high speed electro-optical modulators \cite{Liu2011d,Liu2012d}, photodetectors \cite{Mueller2010,Xia2009a}, saturable absorbers \cite{Bao2009,Sun2010}, and nonlinear media for four-wave mixing \cite{Hendry2010,Zhang2011b,Gu2012}. Intrinsic graphene exhibits absorption of 2.3\% \cite{Mak2008b} in the infrared to visible spectra range. While this absorption coefficient is remarkably high for a single atomic layer, for practical applications, a larger absorption coefficient is needed. To increase the light-matter interactions in graphene, approaches to date have included the integration of graphene with optical micro-cavities \cite{Furchia,Majumdar2013,Engel2012,Gan2014,Gan2013a}, plasmonic nanostructures \cite{Echtermeyer2011,Fang2012,Fang2012,Liu2011b,Chen2012l,Fei2012} and silicon photonic waveguides \cite{Liu2011d,Liu2012d,Koester2012,Li2012e}.

In this paper, we demonstrate a graphene photodetector integrated in a linear defect cavity defined in a planar photonic crystal (PPC). A single graphene layer strongly couples to the cavity evanescent field, increasing the light-matter interaction in graphene \cite{Gan2012} for photocurrent generation. Coupled mode theory predicts maximal absorption into the graphene absorber when the intrinsic cavity loss rate, $\kappa_c$, equals the loss rate into the graphene sheet, $\kappa_{cg}$ \cite{Gan2012}. Upon optimization of the cavity design, we obtain nearly critical coupling with $\kappa_{cg}/\kappa_c\approx1.3$ and observe an eight-fold enhancement of photocurrent in the graphene photo-detector. The observed reflectivity and photocurrent spectra in the graphene detector agree well with the coupled graphene-cavity model. Spatial mapping of the photocurrent allows us to compare the response of the graphene detector with and without optical enhancement via the PPC cavity.

\begin{figure}
\includegraphics[width=17.5cm]{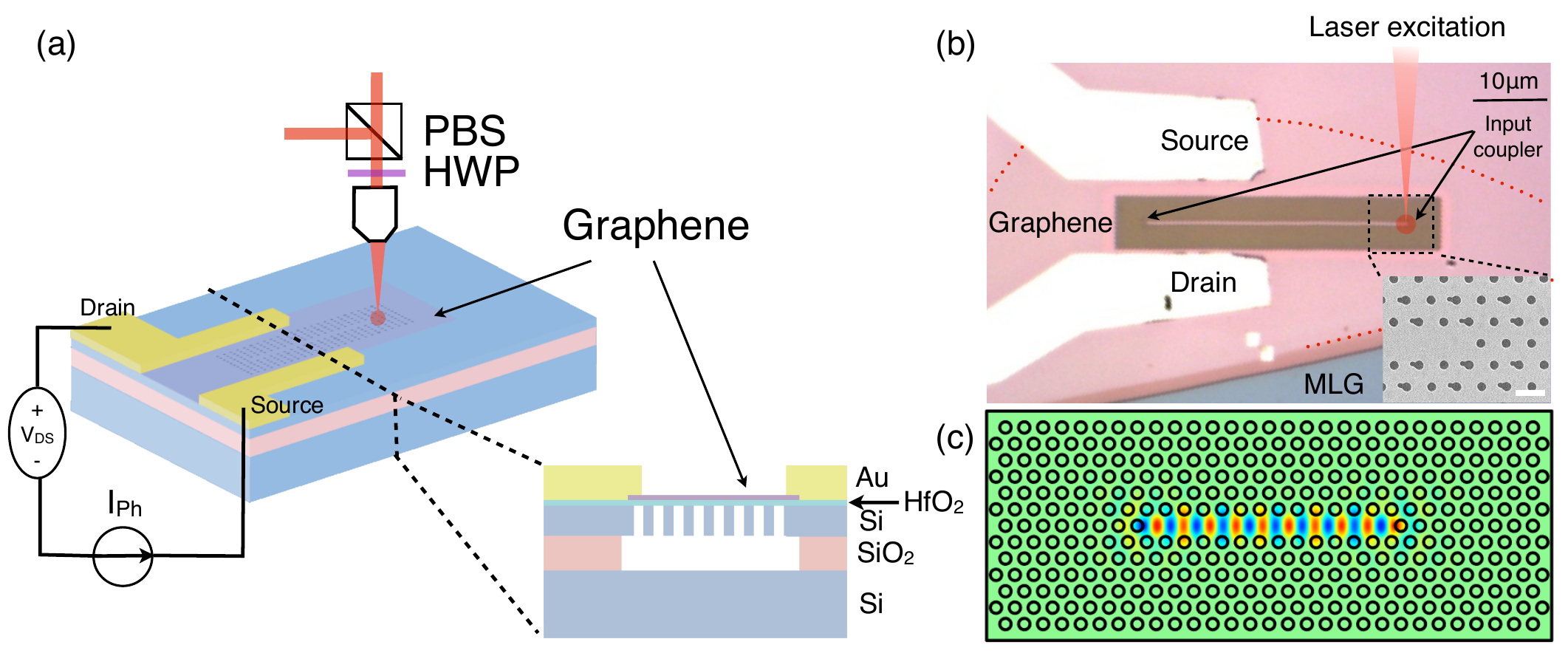}% Here is how to import EPS art
\caption{\label{fig:epsart1}(a) Schematic of the PPC cavity integrated graphene photodetector. (b) Optical image of the graphene detector. A single layer graphene covers the silicon surface in the red dotted region while a multi-layer graphene (MLG) is shown in the blue area. Two metal leads (source and drain) are deposited to contact graphene electrically. Inset: SEM image of one end of the PPC cavity. The enlarged air holes at a period of 2{\it{a}} form a grating in the PPC lattice for optimizing the coupling between graphene, cavity and vertical incident light. Scale bar: 500 nm. (c) FDTD simulation shows a localized resonant mode inside the PPC cavity.}
\end{figure}

The PPC cavity-integrated graphene photodetector is illustrated in Fig. 1(a). An air-suspended PPC cavity was fabricated on a silicon-on-insulator wafer with a 260 nm thick silicon (Si) membrane, using a combination of electron beam lithography (EBL) and dry/wet etching steps. The PPC has a lattice spacing of {\it{a}} = 450 nm and hole radius of 0.29{\it{a}}. A linear defect in the center of the PPC lattice forms a long cavity (Fig. 1(b)), producing bounded cavity modes. A layer of 20 nm hafnium oxide (HfO$_{2}$), deposited by atomic layer deposition (ALD), electrically isolates the metal contacts of graphene from the Si layer. Monolayer graphene was prepared by mechanical exfoliation and then transferred onto the PPC cavity with a precision alignment technique \cite{Dean2010b}. The source and drain contacts were defined by EBL, Ti/Pd/Au (1/20/50 nm) deposition, and lift-off. Fig. 1(b) shows the completed device.

Numerical simulation of the optical field in the long cavity shows strongly localized optical field, as displayed in Fig. 1(c). At the end of the line defects (inset of Fig. 1(b)), a series of perturbations in the PPC lattice at a spatial frequency of $k_x=\pi/2a$ serve to scatter the light vertically upward~ \cite{2011.OpEx.Tsai-Englund.PC-coupler}. This additional loss is also used to match the extrinsic and internal photon loss rates to approach the critical coupling regime of the graphene-cavity system \cite{Gan2012}, as discussed below.

We characterized the PPC cavity using a vertical cross-polarization confocal microscope with a broad-band (super-continuum laser) excitation source, as illustrated in Fig. 1(a). Vertical incident light was coupled at 45$^\circ$ to the cavity polarization and collected at -45$^\circ$ to minimize background light reflected without coupling into the polarized cavity modes\cite{2007.Nature1_etal}. The reflection was analyzed using a commercial spectrometer with a resolution of 0.05 nm. Fig. 2(a) shows the reflection spectrum of the PPC cavity before (blue) and after (green) graphene deposition. Multiple peaks at a wavelength range between 1520 nm and 1550 nm correspond to the resonant modes within the PPC bandgap. After graphene is transferred onto the cavity, the resonant peaks are lowered and broadened, as expected for excess loss in the cavity due to graphene absorption. For the multi-mode cavity used in this experiment, the intensity reflection coefficient R($\omega$) can be obtained quantitatively from coupled mode for an ensemble of cavity modes  coupled to a common waveguide mode\footnote{See supplementary material at [http://] for the theoretical model of the coupled graphene-cavity system and the extraction of coupling efficiency $\eta$},

%\begin{equation}
%A(\omega)=\frac{\eta\kappa_{c}\kappa_{cg}}{(\omega_{0}+\Delta\omega-\omega)^2+(\kappa_{c}/2+\kappa_{cg}/2)^{2}}
%\end{equation}

\begin{equation}
R(\omega)=\sum_{j}\frac{\eta_j^2\kappa_{cj}^2}{(\omega_{0j}+\Delta\omega_j-\omega)^2+(\kappa_{cj}/2+\kappa_{cgj}/2)^2} 
\end{equation}

\noindent where $\kappa_{cj}$ denote the intrinsic cavity decay rates of mode $j$ and $\eta_j$ are the coupling efficiencies between these cavity modes and the approximately Gaussian modes of the microscope objective. Graphene induces additional cavity loss rates $\kappa_{cgj}$ and cavity resonant frequency shifts $\Delta\omega_j$. With $\Delta\omega_j=0$ and $\kappa_{cgi}=0$, we experimentally extract $\omega_{0j}$, $\kappa_{cj}$ and $\eta_j$ for different modes by fitting the cavity reflection spectra prior to loading with graphene. The fit by Eq. (2) is displayed as the red curve in the top panel of Fig. 2(b), showing good agreement with the six peaks (blue) measured experimentally. Using the same approach, we extract $\kappa_{cgj}$ and $\Delta\omega_j$ from the reflection of the cavity after coupled to graphene and plot the fitting curve in the bottom of Fig. 2(b) (red). Comparing to the measured reflection (green), the fitting shows good agreement from 1522 nm to 1541 nm.

\begin{figure}
\includegraphics[width=14.0cm]{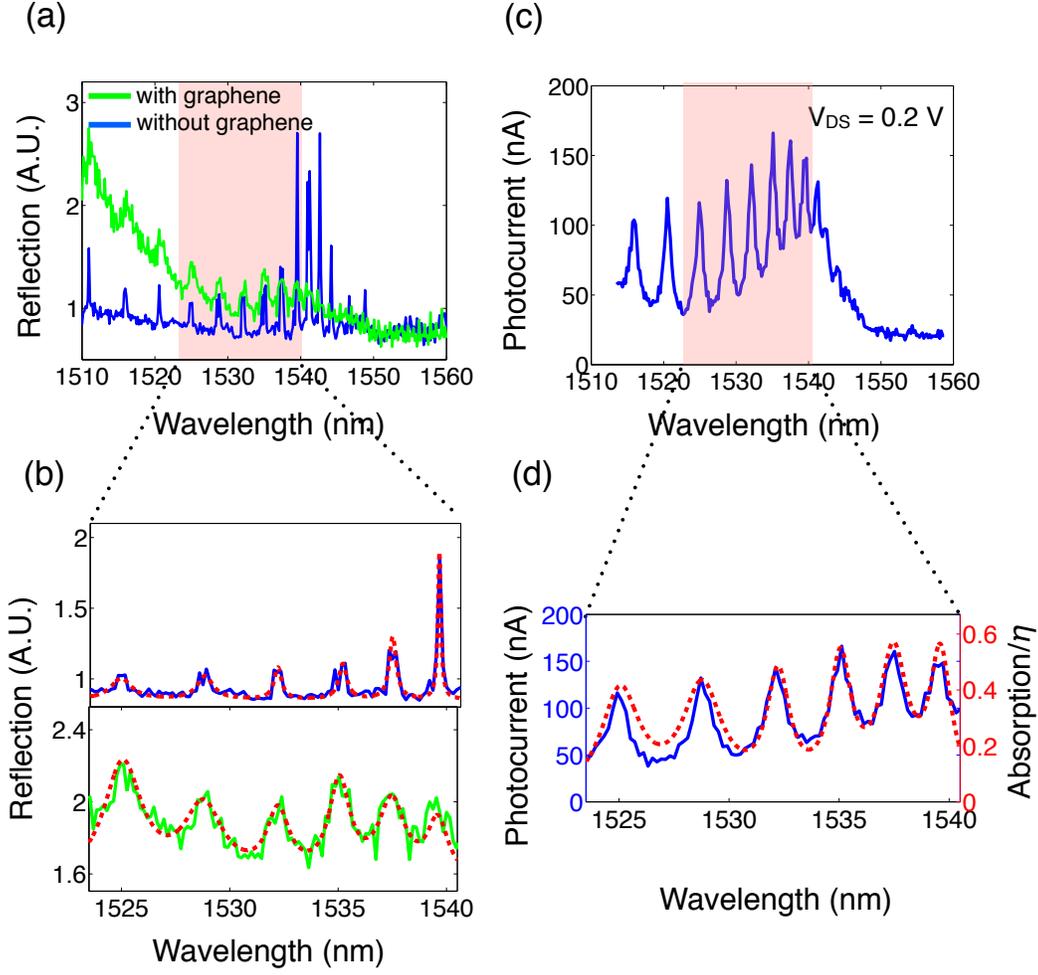}% Here is how to import EPS art
\caption{\label{fig:epsart2}(a) Reflection spectra of the PPC cavity before (blue) and after (green) graphene is deposited on the cavity. (b) A zoom-in view of Fig. 2(a) in the spectral range of 1522 nm to 1541 nm. The red curve in the top panel shows the fitting of the reflection data (blue) measured experimentally. The red and green curves in the bottom panel show the calculated and measured cavity reflection after loading with graphene. (c) The photocurrent spectra of the graphene photodetector measured at the location marked by the red circle in Fig. 1(b) with a bias voltage V$_{DS} = 0.2$ V. (d) Photocurrent spectra (blue curve) of the graphene detector between 1520 nm to 1540 nm, consistent with the absorption spectra of graphene (red curve) derived from coupled mode theory.}
\end{figure}

The photocurrent of the graphene photodetector was measured using a tunable narrowband laser source (Anritsu MG9638) focused onto the sample with a spot size of 1 $\upmu$m. The CW laser is modulated at 20 kHz and the photcurrent is recorded on a lock-in amplifier (SR830) after a current pre-amplifier (SR570). All measurements are performed under ambient conditions. We measure the photocurrent at the location marked by the red circle in Fig. 1(b) while sweeping the incident wavelength from 1520 nm to 1555 nm to obtain the photocurrent spectra. The drain-source bias (V$_{DS}$) was kept at 0.2 V and the input power was 250 $\upmu$W. As displayed in Fig. 2(c), the photocurrent shows multiple spectral peaks, overlapping with the resonant peaks observed in the cavity reflection. For wavelengths below 1550 nm, the input light is enhanced in the PPC line defect, increasing the absorption and correspondingly the photocurrent in graphene. For wavelengths above 1550 nm, the photocurrent drops to a uniformly small value. In this regime, the incident light is detuned from any cavity modes. We attribute the residual photocurrent seen in Fig. 2(c) to the scattering and in-plane guiding of non-resonant light. Comparing the photocurrent when the incident light is on and off a cavity resonance, we observe an up to eight-fold enhancement of the photocurrent at the wavelength of 1535 nm. Using the coupled graphene-cavity model described above, the absorption coefficient into graphene (the fraction of vertically incident light that is ultimately absorbed in graphene) can be expressed as \footnotemark[1]

\begin{equation}
A_g(\omega)=\sum_{j}\frac{\eta_j\kappa_{cj}\kappa_{cgj}}{(\omega_{0j}+\Delta\omega-\omega)^2+(\kappa_{cj}/2+\kappa_{cgj}/2)^2} 
\end{equation}

\noindent We can now deduce the absorption of graphene as a function of input wavelength using the parameters extracted from the reflection curves in Fig. 2(a) and 2(b). As shown in Fig. 2(d), the absorption of graphene normalized by the microscope-cavity coupling efficiency $\eta$ (red) is plotted with the measured photocurrent in graphene (blue). The overlap between the two curves indicates that the photocurrent enhancement originates from the enhanced absorption of graphene in the PPC cavity. The graphene detector operates with a broad bandwidth over the entire cavity modes in the photonic band gap. As the wavelength approaches the band edge of the photonic crystal, the free spectral range of the resonant peaks become smaller, resulting overlapping of resonant peaks. Therefore, the photocurrent is enhanced over a broad spectral range of about 10 nm, as shown in Fig. 2(c).

A spatial mapping of the photocurrent in Fig. 3 elucidates the coupling mechanism into the cavity modes. Figure 3(b-d) map the photocurrent at the locations indicated in the SEM image of Fig. 3(a) with an input excitation wavelength of 1535 nm and bias voltage (V$_{DS}$) of 0.2 V. In the region where graphene is contacted by two metal electrodes, electron-hole pairs are generated by single-pass absorption of the vertically incident beam and then separated by the local electric field. Therefore photocurrent is generated in the channel area shown in Fig. 3(b) and left half of Fig. 3(c). Since the metal leads cover only approximately 50\% of the graphene sheet on the PPC cavity, half of the graphene sheet has nearly zero photocurrent when laser excites these areas, as shown in the right half of Fig. 3(c) and Fig. 3(d). However, the bright spot at the end of the cavity defect in Fig. 3(d) indicates that photocurrent is generated when light couples into the PPC cavity through the input coupler. In this graphene-cavity system, as indicated in Eq. (2), the maximum absorption into graphene occurs when $\kappa_c/\kappa_{cg}$ = 1 with a value of $\eta$. Therefore, the cavity design was optimized by introducing additional loss via the directional couplers at the ends of the cavity to match $\kappa_c$ and $\kappa_{cg}$ while increasing $\eta$. The ratio of $\kappa_c/\kappa_{cg}$ for the device described in this study is 1.3 and the coupling efficiency $\eta$ is 0.04 \footnotemark[1]. Because of low vertical coupling efficiency, the responsivity of our device is 0.6 mA/W, corresponding to an internal quantum efficiency of 0.35 \%. We plot two traces from Fig. 3(b) (red) and 3(d) (magenta) in Fig. 3(e) to compare the photocurrent due to single-pass absorption of graphene and its enhancement after coupling to the cavity. The trace across the PPC cavity coupler in Fig. 3(d) is normalized to $\eta$. As shown in Fig. 3(e), the photocurrent enhancement can reach a maximum factor of 25 when the coupling efficiency is maximized. In practical devices, the coupling efficiency $\eta$ can exceed 45\% with an on-chip edge coupler or tapered fiber coupling \cite{Akahane2005}, indicating that overall efficient light detection is possible. It is important to note that in our device, the photocurrent is generated in the middle of the graphene channel, above the cavity line defect. However, as observed in Fig. 3(a), the photocurrent exhibits stronger response in the vicinity of the metal electrodes. We attribute this enhancement to the built-in electric potential introduced by the doping of the  metal on graphene~ \cite{Mueller2009,Xia2009b}. This indicates that it will be possible to further improve the performance of this graphene detector by placing the metal electrodes closer to the edge of the cavity.

In conclusion, we have demonstrated up to eight-fold enhancement of photocurrent in a graphene photodetector by coupling of the graphene absorber to a photonic crystal cavity. Compared to single-pass absorption, we estimate that if light were efficiently coupled into the defect state, i.e. $\eta\sim 1$, the absorption efficiency could reach as high as $95$\% with $\kappa_c/\kappa_{cg}=1.3$. Coupling efficiency from waveguides into photonic crystal cavities has been demonstrated previously with near unity efficiency\cite{Takano:06,Lin2010b}. At the Brillouin zone boundary of the PPC, the cavity resonant modes overlap and span a broad band (10 nm) of enhanced absorption and photocurrent in graphene. The photocurrent shows good agreement with the calculated absorption spectra from the optical reflection data based on a coupled graphene and cavity model. While we have not performed high-speed measurements, graphene photodetctors have already been shown to enable high-speed optical communication \cite{Mueller2010,Xia2009a,Gan2013}. The PPC-cavity-coupled graphene device demonstrated here shows the feasibility of efficient and ultra-compact graphene photodetectors in a chip-integrated architecture.

\begin{figure}
\includegraphics[width=11.0cm]{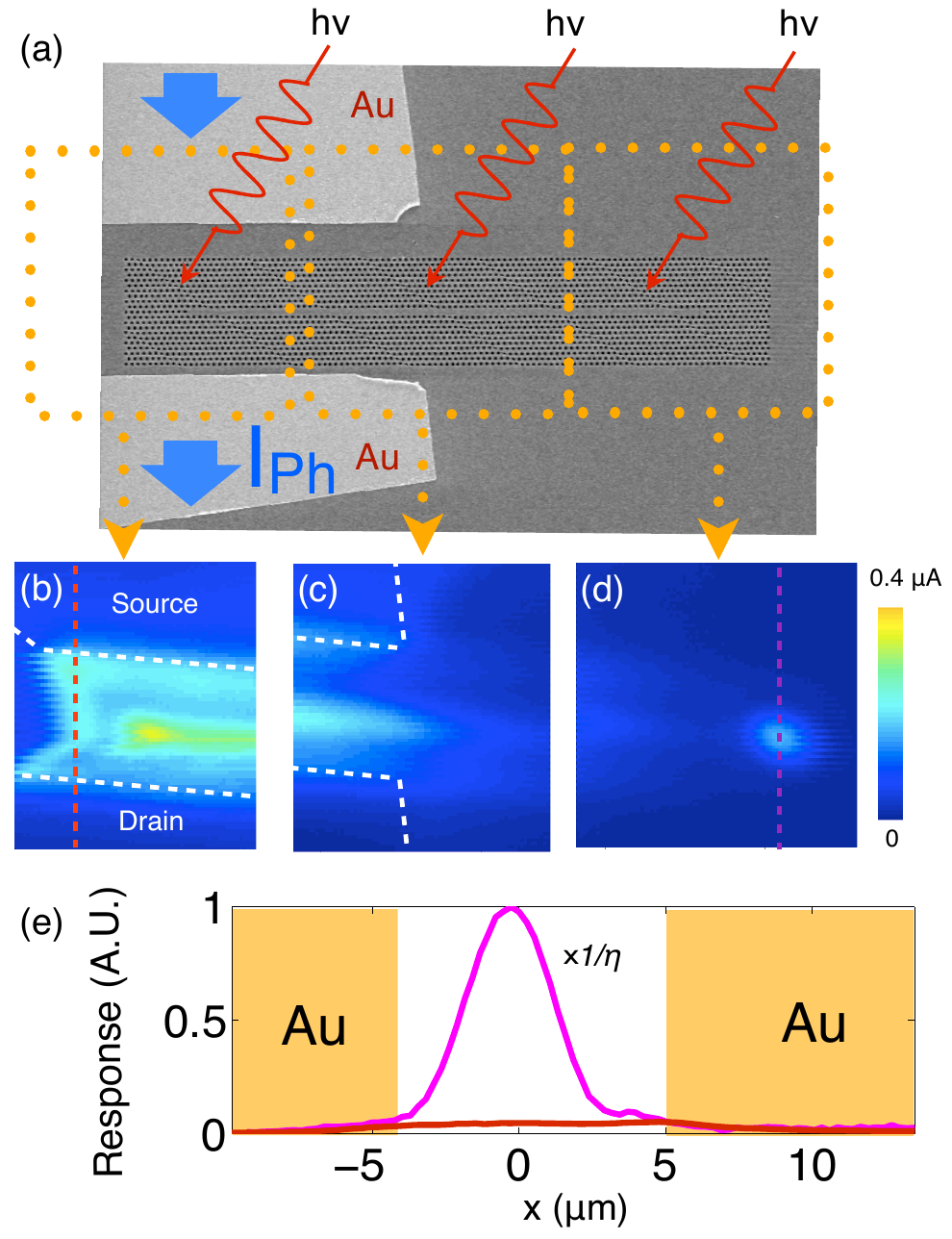}% Here is how to import EPS art
\caption{\label{fig:epsart3}(a) SEM image of the graphene detector on the PPC cavity. Scale bar: 3 $\upmu$m. (b-d) Spatial mapping of the photocurrent at the three areas indicated in (a). The white dashed lines show the boundary of the metal contacts. (e) The trace of photocurrent data in (b) and (d), showing the comparison of the photocurrent with (magenta) and without (red) coupled to the PPC cavity. The trace across the coupler of the PPC cavity is normalized to the coupling efficiency $\eta$, and the edge of the metal are labeled by the yellow shaded region.}
\end{figure}

\begin{acknowledgments}

This work was supported in part by the Center for Excitonics, an Energy Frontier Research Center funded by the U.S. Department of Energy, Office of Science, Office of Basic Energy Sciences under Award Number DE-SC0001088. Device fabrication was carried out at the Center for Functional Nanomaterials, Brookhaven National Laboratory, which is supported by the U.S. Department of Energy, Office of Basic Energy Sciences, under Contract No. DE-AC02-98CH10886, and at WPAFB, Ohio. Graphene assembly was supported by the Center for Re-Defining Photovoltaic Efficiency Through Molecule Scale Control, an Energy Frontier Research Center funded by the U.S. Department of Energy, Office of Science, Office of Basic Energy Science under Award Number DE-SC0001085. The authors thank Nicholas Harris for the helpful discussions.

\end{acknowledgments}

\bibliography{library}% Produces the bibliography via BibTeX.

\end{document}